\documentclass{llncs}
\usepackage[latin1]{inputenc}
\usepackage[T1]{fontenc}
\usepackage{graphicx}
\usepackage{amssymb}
\usepackage{amsfonts}
\usepackage{amsxtra}

\title{Probabilistic Nets-within-Nets}

\author{
 Michael K\"ohler-Bu\ss{}meier\orcidID{0000-0002-3074-4145}
}

\institute{%
  University of Applied Science Hamburg 
 \\  Berliner Tor 7, D-20099 Hamburg, Germany
  \\  \email{michael.koehler-bussmeier@haw-hamburg.de} 
}

\newcommand{\prXOR}[2]{
{}^{\langle #1\rangle}{\oplus}{}^{\langle #2\rangle}
}
\newcommand{\color}[2][1]{}

\newcommand{\Sonar}{\textsc{Sonar}}
\newcommand{\Hornet}{\textsc{Hornet}}
\newcommand{\Hornets}{\textsc{Hornets}}

\newcommand{\eHornet}{\textsc{eHornet}}
\newcommand{\eHornets}{\textsc{eHornets}}
\newcommand{\EOS}{\textsc{Eos}}
\newcommand{\skipit}[1]{}
\newcommand{\postit}[1]{\marginpar{{\parbox{\linewidth}{\begin{center}\footnotesize\textsf{#1}\end{center}}}}}
\renewcommand{\postit}[1]{}
\newcommand{\ms}[1]{\mathbf{#1}}

\newcommand{\labtr}[2][{}]{\xrightarrow[#1]{#2}}
\newcommand{\labtra}[2]{\xrightarrow[#2]{#1}}

\newcommand{\Nat}{\mathbb{N}} 
 
\newcommand{\Real}{\mathbb{R}}

\newcommand{\sn}[1]{\widehat{#1}} 
\newcommand{\on}[1]{#1}
 \newcommand{\postset}[1]{{{#1}^\bullet}}
 \newcommand{\preset}[1]{{{}^\bullet #1}}

\newcommand{\pre}{\mathbf{pre}} 
\newcommand{\post}{\mathbf{post}}

\newcommand{\key}[1]{\emph{#1}}

\begin{document}
\maketitle

\begin{abstract}
  In this paper we study \Hornets{} extended with firing
  probabilities.
  \Hornets{} are a Nets-within-Nets formalism, i.e., a Petri net
  formalism where the tokens are Petri nets again.  Each of these
  net-tokens has its own firing rate, independent from the rates of
  other net-tokens.

  \Hornets{} provide algebraic operations to modify net-tokens during
  the firing. For our stochastic extension these operators could also
  modify the net-token's firing rate.

  We use our model to analyse self-modifying systems quantitatively.
  \Hornets{} are very well suited to model self-adaptive systems
  performing a MAPE-like loop (monitoring-analyse-plan-execute).
  Here, the system net describes the loop, and the net-tokens describe
  the adapted model elements.

\end{abstract}

\section{Introduction}

\Hornets{} are a well-suited formalism to specify self-modification as
it has inbuilt constructs to support structural modifications of Petri
nets as an effect transitions firing.
We use \Hornets{} to model the self-adaption processes in multi-agent
system (cf. \cite{Sudeikat+22,Koehler23-mape-micmac}) specified in
\Sonar{} \cite{koehler+2009-topnoc} -- forming a so-called MAPE-K-loop
(short for: monitoring-analyse-plan-execute-knowledge
\cite{Weyns2020}).
In this application area the adaptation also includes structural modifications of the
Petri net-token, at run-time; it is the  system's architecture, which is dynamic.

\Hornets{} \cite{koehler09-hornets} follow the \emph{Nets-within-Nets}
approach \cite{Valk-acpn}, i.e., we have Petri nets that have
\emph{nets as tokens} and we have algebraic operations on the
net-tokens.
A net-token is a pair $[\on{N}, \on{m}]$, where $\on{N}$ is the
object-net defining the topology and $\on{m}$ is the current marking
of the net-token.  Firing transforms the net-token's marking $\on{m}$
and the algebraic operators modify the net topology $\on{N}$.

In this paper, we like to extend our
approach by quantitative information about the
relative frequencies of self-modifications.
Our main contribution is the definition of \key{Stochastic
  \Hornets{}}, where object-nets have the form $\on{N}^{\Lambda}$; the
mapping $\Lambda$ assigns firing rates to the object net's
transitions, which induce firing probabilities (cf.~our example in
Section~\ref{sec:stoch-ehornet-bos}).

We like to mention two special features: Firstly, we have an
independent rate for each net-token $[\on{N}^{\Lambda}, \on{m}]$.
Secondly, for \Hornets{} we could use the operators to modify  the firing rates $\Lambda$.
We exploit this feature during the monitoring phase of our self-adaptation loop.
From the two specialties we obtain that the state space is more
complex to analyse when compared to e.g. Stochastic Nets
\cite{marsan1989}.

\paragraph{Related Work}

Probabilistic choices in process models go back to the work on
probabilistic automata \cite{RABIN1963230}.
For Petri nets probability is heavily used to model the firing time
distribution, like for Stochastic Nets \cite{marsan1989}; so, the
concept is more related to `time' than to `alternatives'.  However,
probabilistic choices are introduced by immediate transitions and
their rates for Generalised Stochastic Petri Nets (GSPN)
\cite{modelling-gspn-95}.
Probabilistic Choices for Process Algebra are studied in \cite{10.1007/BFb0084810}.
These formalisms are also used for adaptive systems, e.g.,
\cite{10.1145/2330667.2330686} argues that self-adaptive software
needs verification at run-time using model-checking \cite{KNP11}.

However, these formalism uses `flat' models; here we consider nested
structures, where execution is embedded into a context, like the
Ambient Calculus \cite{Cardelli+99b} and the $\pi$-calculus
\cite{Milner92a} (the process algebra perspective) or nets-with-nets
\cite{Valk-acpn} formalisms, like nested nets \cite{Lomazova00},
elementary object systems (\EOS{}) \cite{Koehler14-fi-lam}, and
\Hornets{} \cite{koehler09-hornets}.

Object Nets can be seen as the Petri net perspective on mobility, in
contrast to the Ambient Calculus \cite{Cardelli+99b} or the
$\pi$-calculus \cite{Milner92a}, which form the process algebra
perspective.
While probabilistic extensions exist for context-oriented process
algebras (cf. \cite{KWIATKOWSKA20091272} and\cite{PRADALIER2006119}),
to the best of our knowledge there are no such extensions for
nets-within-nets.

\paragraph{Structure}
The paper has the following structure.
In Section~\ref{sec:hornets-intro} we recall the definition of
\Hornets{} as given in previous publications.  In
Section~\ref{sec:stochastic-ehornets} we present the main contribution
of this paper, namely Stochastic \eHornets{}.
In Section~\ref{sec:stochastic-ehornets-for-mape} we show how we use
this formalism to model and analyse self-adaptive systems and how the
firing rates of the system-net are connected to the transformation
complexity.
We present an example for a MAPE-like adaption in
Section~\ref{sec:stoch-ehornet-bos}, where we study the well-known
\textit{battle-of-sexes} coordination game.
The work ends with a conclusion and an outlook to ongoing work.

\section{Algebraic Nets-within-Nets: Hornets }
\label{sec:hornets-intro}

We have defined \Hornets{} in \cite{koehler09-hornets} as a
generalisation of our object nets \cite{koehler+09-fi}, which follow
the \emph{nets-within-nets} paradigm as proposed by Valk
\cite{Valk-acpn}.  


In the following we will present the simplified model of \key{Elementary
Hornets} (\eHornets{}) from \cite{Koehler13-fi-hornets}, where the
nesting strcuture is restricted to two levels, while \Hornets{}
\cite{koehler09-hornets} allow for an arbitrarily nested structure.
This is done in analogy to the class of \emph{elementary object net
  systems} (\textsc{Eos}) \cite{koehler+09-fi}, which are the
two-level specialisation of general object nets \cite{koehler+09-fi}.

\begin{figure}[htbp]  \centerline{\includegraphics[width=0.99\textwidth]{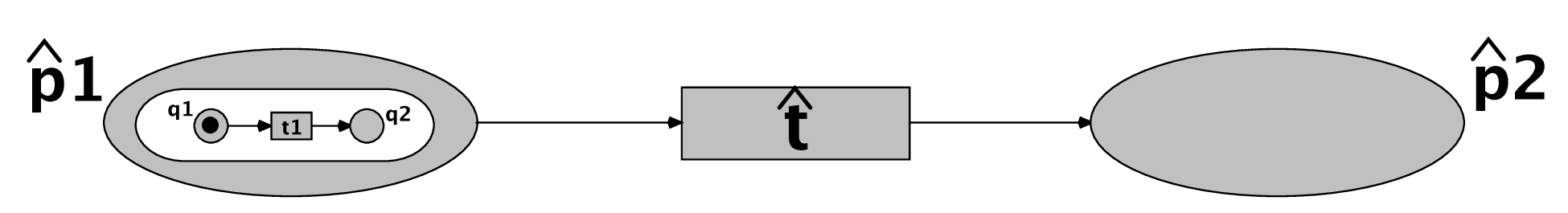}}
  \caption{\label{fig:eos} Nets within Nets: Nets as Tokens}
\end{figure}%

\begin{example}  
  With \Hornets{} we study Petri nets where the tokens are nets again,
  i.e., we have a nested marking.
  Assume that we have the object net $\on{N} $ with places
  $\on{P} = \{ \on{q}_1, \on{q}_2\}$ and transitions
  $\on{T} = \{ \on{t}_1\}$.
  The marking of the \Hornet{} of Figure~\ref{fig:eos} is denoted by
  the nested multiset:
  \(
  \sn{p}_1[
  \on{N},
  \on{q}_1
  ]
  \).
  Events are also nested.  We have three
  different kinds of events -- as illustrated by the example
  given in Figure~\ref{fig:eos}:
  \begin{enumerate}
  \item System-autonomous: The system net transition $\sn{t}$ fires
    autonomously, which moves the net-token from $\sn{p}_1$ to $\sn{p}_2$
    without changing its marking.
    \[
      \sn{p}_1[
      \on{N},
      \on{q}_1
      ]
      \quad
      \to
      \quad
      \sn{p}_2[
      \on{N},
      \on{q}_1
      ]
    \]

  \item Object{} autonomous: The object{} net fires transition $\on{t}_1$
    ``moving'' the black token from $\on{q}_1$ to $\on{q}_2$.  The object{} net
    remains at its location $\sn{p}_1$.
    \[
      \sn{p}_1[
      \on{N},
      \on{q}_1
      ]
      \quad
      \to
      \quad
      \sn{p}_1[
      \on{N},
      \on{q}_2
      ]
    \]

  \item Synchronisation: Whenever we add matching synchronisation
    inscriptions (using communication channels) at the system net
    transition $\sn{t}$ and the object net transition $\on{t}_1$, then
    both must fire synchronously: The object{} net is moved to
    $\sn{p}_2$ and the black token moves from $\on{q}_1$ to $\on{q}_2$
    inside.
    Whenever synchronisation is specified, autonomous actions are
    forbidden.

  \end{enumerate}
\end{example}

\smallskip

For \Hornets{} we extend object{} nets with algebraic concepts
that allow to modify the structure of the net-tokens as a result of a
firing transition.  This is a generalisation of the approach of
algebraic nets \cite{Reisig:91}, where algebraic data types replace
the anonymous black tokens.

It is not hard to prove that the general \Hornet{} formalism is
Turing-complete: In \cite{koehler09-hornets} we have proven that
there are several possibilities to simulate counter programs: One
could use the nesting to encode counters. Another possibility is to
encode counters in the algebraic structure of the net operators.


In the following we recall the definition of \eHornets{} from
\cite{Koehler13-fi-hornets}. 
First, we recall notations for p/t nets; 
then we will define the algebraic structure of the net-token and the logic used for guards.
We introduce nested multisets as the marking structure.
Finally, we  define the firing rule, that, in general, 
involves a synchronisation of system net transitions  
with transitions of the net-tokens, i.e., we define nested events.
The reader familiar with the firing rule of \eHornets{}
(Def.~\ref{def:ehornet-fire})  can safely skip the remainder of
this section.

\paragraph{Multisets and P/T Nets}

A multiset $\ms{m}$ on the set $D$ is a mapping $\ms{m}: D \to \Nat$.
Multisets can also be represented as a formal sum in the form
$\ms{m} = \sum_{i=1}^{n} x_i$, where $x_i \in D$.

Multiset addition
is defined
component-wise: $(\ms{m}_1 + \ms{m}_2)(d) := \ms{m}_1(d) +
\ms{m}_2(d)$.
The empty multiset $\mathbf{0}$ is defined as $\mathbf{0}(d)
= 0$ for all $d \in D$.  
Multiset-difference $\ms{m}_1 - \ms{m}_2$ is defined by $(\ms{m}_1 -
\ms{m}_2)(d) := \max(\ms{m}_1(d) - \ms{m}_2(d), 0)$.

The cardinality of a multiset is $|\ms{m}| := \sum_{d \in D}
\ms{m}(d)$.
A multiset $\ms{m}$ is finite if $|\ms{m}| < \infty$.  The set of all
finite multisets over the set $D$ is denoted $\mathit{MS}(D)$.
The \key{domain} of a multiset is
$\mathit{dom}(\ms{m}) := \{ d \in D \mid \ms{m}(d) > 0 \}$.

Multiset notations are used for sets as well. The meaning will be
apparent from its use.

Any mapping $f: D \to D'$ extends to a multiset-homomorphism
$f^\sharp: \mathit{MS}(D) \to \mathit{MS}(D')$ by \(
f^\sharp\left(\sum_{i=1}^{n} x_i\right) = \sum_{i=1}^n f(x_i) \).

A \key{p/t net} $N$ is a tuple \( N = (P, T, \pre,\post), \) such
that $P$ is a set of places, $T$ is a set of transitions, with $P
\cap T = \emptyset$, and $\pre, \post: T \to \mathit{MS}(P)$ are the
pre- and post-condition functions.  A marking of $N$ is a multiset
of places: $\ms{m} \in \mathit{MS}(P)$.  
We denote the enabling of $t$ in marking
$\ms{m}$ by $\ms{m} \labtra{t}{}$. Firing of $t$ is denoted by
$\ms{m}\labtra{t}{} \ms{m}'$.


\paragraph{Net-Algebras}

We define the algebraic structure of object{} nets.  For a general
introduction of algebraic specifications cf.  \cite{EhrigMahr:92}.


Let $K$ be a set of net-types (kinds).  A (many-sorted)
\key{specification} $(\Sigma, X, E)$ consists of a signature $\Sigma$,
a family of variables $X = (X_k)_{k \in K}$, and a family of axioms $E
= (E_k)_{k \in K}$.

A \key{signature} is a disjoint family $\Sigma = (\Sigma_{k_1\cdots
  k_n,k})_{k_1,\cdots, k_n,k \in K}$ of operators.
The set of terms of type $k$ over a signature $\Sigma$ and variables
$X$ is denoted $\mathbb{T}^{k}_{\Sigma}(X)$.

We use (many-sorted) predicate logic, where the terms are generated by
a signature $\Sigma$ and formulae are defined by a family of
predicates $\Psi = (\Psi_{n})_{n \in \Nat}$.  The set of formulae is
denoted $\mathit{PL}_{\Gamma}$, where $\Gamma = (\Sigma,X,E,\Psi)$ is
the \key{logic structure}.

\paragraph{Object Nets and Net-Algebras}

Let $\Sigma$ be a signature over $K$.  A \key{net-algebra} assigns to
each type $k \in K$ a set $\;\mathcal{U}_k$ of object{} nets -- the net
universe.
Each object $\on{N} \in \mathcal{U}_k, k \in K$ net is a p/t
net $\on{N} = (\on{P}_{\on{N}}, \on{T}_{\on{N}}, \pre_{\on{N}},
\post_{\on{N}})$.
We identify $\mathcal{U}$ with $\bigcup_{k \in K} \mathcal{U}_{k}$ in the
following.  
We assume the family $\mathcal{U} = (\mathcal{U}_k)_{k \in K}$ to be disjoint.

The nodes of the object{} nets in $\mathcal{U}_k$ are not disjoint, since the
firing rule allows to transfer tokens between net tokens within the
same set $\mathcal{U}_k$.  Such a transfer is possible, if we assume that
all nets $N \in \mathcal{U}_k$ have the same set of places $\on{P}_k$.  
$\on{P}_k$ is the place universe for all object{} nets of kind $k$.

The family of object{} nets $\mathcal{U}$ is the universe of the algebra.  A
\key{net-algebra} $(\mathcal{U}, \mathcal{I})$ assigns to each constant $\sigma
\in \Sigma_{\lambda,k}$ an object{} net $\sigma^{\mathcal{I}} \in
\mathcal{U}_k$ and to each operator $\sigma \in \Sigma_{k_1\cdots k_n,k}$
with $n >0$ a mapping $\sigma^\mathcal{I}: (\mathcal{U}_{k_1} \times \cdots
\times \mathcal{U}_{k_n}) \to \mathcal{U}_k$.

A variable assignment $\alpha = (\alpha_k : X_k \to \mathcal{U}_k)_{k \in
  K}$ maps each variable onto an element of the algebra.  For a
variable assignment $\alpha$ the evaluation of a term $t \in
\mathbb{T}_{\Sigma}^k(X)$ is uniquely defined and will be denoted as
$\alpha(t)$.

A net-algebra, such that all axioms of $(\Sigma, X, E)$ are valid, is called \emph{net-theory}.

\paragraph{Finite Models}

In general, $\on{P}_k$ is not finite. Since we like each object{} net
to be finite in some sense, we require that the transitions
$\on{T}_{\on{N}}$ of each $\on{N} \in \mathcal{U}_k$ use only a finite
subset of $\on{P}_k$, i.e.,
$\forall \on{N} \in \mathcal{U}: |\preset{\on{T}_{\on{N}}} \cup
\postset{\on{T}_{\on{N}}}| < \infty$.

A net-algebra is called \key{finite} if $\on{P}_k$ is a finite set for
each $k \in K$.

Since all nets $N \in \mathcal{U}_k$ have the same set of places
$\on{P}_k$, which is required to be finite for \eHornets{}, there is
an upper bound for the cardinality of $\mathcal{U}_k$.

\begin{proposition}[Lemma 2.1 in \cite{Koehler13-fi-hornets}]
  \label{lem:bound-of-on}
  For each $k \in K$ the cardinality of each net universe
  $\mathcal{U}_k$ is bound as follows: $|\mathcal{U}_k | \leq
  2^{\left( 2^{4|\on{P}_k|} \right)} $.
\end{proposition}

\subsection{Nested Markings and Synchronisation}

\paragraph{Nested Markings}
A marking of an  \eHornet{} assigns to each system net place
one or many net-tokens.  The places of the system net are typed by the
function \( \mathit{k}: \sn{P} \to K \), meaning that a place $
\sn{p}$ contains net-tokens of kind $\mathit{k}( \sn{p})$.  Since the
net-tokens are instances of object{} nets, a \key{marking} is a
\emph{nested} multiset of the form:
\[
  \mu = \sum_{i=1}^{n} \sn{p}_i[\on{N}_i, \on{M}_i]
  \quad\text{where}\quad
  \sn{p}_i \in \sn{P},
  \on{N}_i \in \mathcal{U}_{\mathit{k}(\sn{p_i})},
  \on{M}_i \in \mathit{MS}(\on{P}_{\on{N}_i}),
  n \in \Nat
\]
Each addend $\sn{p}_i[\on{N}_i, \on{M}_i]$ denotes a net-token on the
place $\sn{p}_i$ that has the structure of the object{} net $\on{N}_i$
and the marking $\on{M}_i \in
\mathit{MS}(\on{P}_{\on{N}_i})$.
The set of all nested multisets is denoted as $\mathcal{M}_H$.
We define the partial order $\sqsubseteq$ on nested multisets by setting
$\mu_1 \sqsubseteq \mu_2$ iff $\exists \mu: \mu_2 = \mu_1 + \mu$.

\paragraph{Projections}

The projection $\Pi^{{1}}_{\on{N}}(\mu)$ is the multiset of all system-net
places that contain the object{} net $\on{N}$:
\begin{equation}
  \Pi^{{1}}_{\on{N}}\left(\sum\nolimits_{i=1}^{n}
    \sn{p}_i[ \on{N}_i, \on{M}_i]   \right)
  := \sum\nolimits_{i=1}^{n} \mathbf{1}_{\on{N}}(\on{N}_i) \cdot \sn{p}_i  
\end{equation}
where the indicator function $\mathbf{1}_{\on{N}}$ is defined as:
$\mathbf{1}_{\on{N}}(\on{N}_i) = 1$ iff $\on{N}_i = \on{N}$.

Analogously, the projection $\Pi^{{2}}_{\on{N}}(\mu)$ is the multiset of all
net-tokens' markings (that belong to the object{} net $\on{N}$):
\begin{equation}
  \label{eq:21}
  \Pi^{{2}}_{\on{N}} \left(\sum\nolimits_{i=1}^{n}  
    \sn{p}_i[\on{N}_i, \on{M}_i]  \right)
  := \sum\nolimits_{i=1}^{n}
  \mathbf{1}_{k}(\on{N}_i) \cdot \on{M}_i  
\end{equation}

The projection $\Pi^{{2}}_{k}(\mu)$ is the sum of all net-tokens' markings
belonging to the same type $k \in K$:
\begin{equation}
  \Pi^{{2}}_{k} \left(\mu \right)
  := \sum\nolimits_{\on{N} \in \mathcal{U}_k}
  \Pi^{{2}}_{\on{N}} \left(\mu\right)  
\end{equation}

\paragraph*{Synchronisation}

The transitions in an \Hornet{} are labelled with synchronisation
inscriptions. We assume a fixed set of channels $C = (C_k)_{k \in K}$.
\begin{itemize}
\item The function family $\sn{l}_{\alpha} = (\sn{l}^{k}_{\alpha})_{k
    \in K}$ defines the synchronisation constraints.
  Each transition of the system net is labelled with a multiset
  $\sn{l}^{k}(\sn{t}) = (e_1,c_1) + \cdots + (e_n,c_n)$, where the
  expression $e_i \in \mathbb{T}^{k}_{\Sigma}(X)$ describes the called
  object{} net and $c_i \in C_k$ is a channel.
  The intention is that $\sn{t}$ fires synchronously with a multiset
  of object{} net transitions with the same multiset of labels.
  (Since we like to express that we synchronise with net-tokens from
  the preset, we usually choose the $e_i$ as one an addend from
  $\pre(\sn{t}) (\sn{p})$ for some $ \sn{p}$, but this is not enforced
  as a formal  restriction.)
  Each variable assignment $\alpha$ generates the function
  $\sn{l}^{k}_{\alpha}(\sn{t})$ defined as:
  \begin{equation}
    \sn{l}^{k}_{\alpha}(\sn{t})(\on{N}) 
    :=
    \sum\nolimits_{1\leq i \leq n \atop \alpha(e_i)=\on{N}}  c_i
    \quad
    \text{for}
    \quad
    \sn{l}^{k}(\sn{t}) =    \sum\nolimits_{1\leq i \leq n}  (e_i,c_i)
  \end{equation}

  Each function $\sn{l}^{k}_{\alpha}(\sn{t})$ assigns to each object
  net $\on{N}$ a multiset of channels.

\item For each $\on{N} \in \mathcal{U}_k$ the  function
  $\on{l}_{\on{N}}$ assigns to each transition $\on{t} \in T_{\on{N}}$
  either a channel $c \in C_k$ or $\bot_{k}$, whenever $\on{t}$
  fires without synchronisation, i.e.,  autonomously.

\end{itemize}

\subsection{Elementary \Hornets{}}

\paragraph*{System Net}

Assume we have a fixed logic $\Gamma = (\Sigma,X,E,\Psi)$ and a
net-theory $(\mathcal{U}, \mathcal{I})$.
An \emph{elementary higher-order object net} (\eHornet{}) is
composed of a system net $\sn{N}$ and the set of object{} nets $\mathcal{U}$.
W.l.o.g. we assume $\sn{N} \not\in \mathcal{U}$.
To guarantee finite algebras for  \eHornets{}, we
require that the net-theory $(\mathcal{U}, \mathcal{I})$ is finite,
i.e., each place universe $\on{P}_k$ is finite.
The system net is a net $\sn{N} = (\sn{P}, \sn{T}, \pre, \post,
\sn{G})$, where each arc is labelled with a multiset of terms:
\(
\pre, \post: \sn{T} \to (\sn{P} \to \mathit{MS}( \mathbb{T}_{\Sigma}(X) ))
\).
Each transition is labelled by a guard predicate
\(
\sn{G} : \sn{T} \to \mathit{PL}_{\Gamma}.
\)
The places of the system net are typed by the function
\(
\mathit{k}: \sn{P} \to K
\).
As a typing constraint we have that each arc inscription has to be a
multiset of terms that are all of the kind that is assigned to the
arc's place:
\begin{equation}
  \pre(\sn{t}) (\sn{p}),\quad
  \post(\sn{t}) (\sn{p}) 
  \quad\in\quad \mathit{MS}(\mathbb{T}^{\mathit{k}(\sn{p})}_{\Sigma}(X) )  
\end{equation}

For each variable binding $\alpha$ we obtain the evaluated functions
$\pre_{\alpha}, \post_{\alpha} : \sn{T} \to (\sn{P} \to \mathit{MS}(
\mathcal{U})) $ in the obvious way.

\begin{definition}[Elementary Hornet, \eHornet]
  \label{def:Hornet}
  Assume a fixed many-sorted predicate logic $\Gamma = (\Sigma, X, E,
  \Psi)$.
  
  An elementary \Hornet{} is a tuple \( \mathit{EH} =
  (\sn{N}, \mathcal{U}, \mathcal{I}, \mathit{k}, l,   \mu_0) \) such that:
  \begin{enumerate}
  \item $\sn{N}$ is an algebraic net, called the \key{system net}.   
  \item $(\mathcal{U}, \mathcal{I})$ is a finite net-theory for the logic $\Gamma$.
  \item $\mathit{k}: \sn{P} \to K$ is the typing of the system net places.
  \item $l = (\sn{l}, \on{l}_{\on{N}})_{ \on{N} \in \mathcal{U}}$ is
    the labelling.

  \item $\mu_{0} \in \mathcal{M}_H$ is the initial marking.
  \end{enumerate}
  
\end{definition}

\begin{example}
  We will illustrate Def.~\ref{def:Hornet} with the example given in
  Figure~\ref{fig:hornet1}.\postit{Use a GSM as Example.}
    We assume that we have one net type: $K = \{  \text{WFN} \}$.
    We have only one operator  $\|$ for parallel composition:
    $\| \in \Sigma_{\text{WFN}^2, \text{WFN}}$.
    The operator is interpreted by $\mathcal{I}$
    as the usual AND operation on workflow nets.    
    We have the universe with three object nets:
    $ \mathcal{U}_{\text{WFN}}= \{ \on{N}_1, \on{N}_2, \on{N}_3 \}$.
    All places of the system net have the same type, i.e.,
    $k( \sn{p} )
    =
    k( \sn{q} )
    =
    k( \sn{r} ) = \text{WFN}$.
    The structure of the system net $\sn{N}$  and the object nets 
    is given the usual way as shown in Fig.~\ref{fig:hornet1}.
    This \eHornet{}  uses no communication channels, i.e., all events occur
    autonomously: $\sn{l}^{k}_{\alpha} = \mathbf{0}$ and
    $\on{l}_{\on{N}} (\on{t}) = \bot_{k}$.
    In the initial marking we consider a \eHornet{} with two nets
  $\on{N}_1$ and $\on{N}_2$ as tokens (as shown on the left):
  \(
      \mu_0 =
      \sn{p}[\on{N}_1, \on{v}]
      +
      \sn{q}[\on{N}_2, \on{s}]
    \).

  \begin{figure}[htbp]
    \setlength{\unitlength}{0.214cm}
    \begin{center}
       \bigskip
      \begin{picture}(53,18)(-4,-7)

        \put(2,1){\circle{2}}
        \put(2,1){\circle*{0.7}}
        \put(3,2){\makebox(1,1){$\sn{p}$}}
        \put(3,1){\vector(2,1){3}}
        \put(4,0.5){\makebox(1,1){$x$}}
        \put(2,5){\circle{2}}
        \put(2,5){\circle*{0.7}}
        \put(3,6){\makebox(1,1)[c]{$\sn{q}$}}
        \put(3,5){\vector(2,-1){3}}
        \put(4,4.5){\makebox(1,1){$y$}}
        \put(6,2){\framebox(2,2)[c]{$\sn{t}$ }}
        \put(8,3){\vector(1,0){4}}
        \put(9,3.3){\makebox(2,1)[c]{$(x \| y)$}}
        \put(13,3){\circle{2}}
        \put(14,3.7){\makebox(1,1)[c]{$\sn{r}$}}

        \put(2,5){\vector(-4,3){4}}
        \put(2,1){\vector(-4,-3){4}}

        \put(-5,-6){
          \setlength{\unitlength}{0.18cm}
          \begin{picture}(37,5)(-3,-1)
            \put(-3,-1){\framebox(37,5)[c]{}}
            \put(-2,2){\makebox(1,1)[c]{$\on{N}_1$}}
            \put(1,1){\circle{2}}
            \put(2,2){\makebox(1,1)[c]{$i_1$}}
            \put(2,1){\vector(1,0){3}}
            \put(5,0){\framebox(2,2)[c]{$a$}}
            \put(7,1){\vector(1,0){3}}
            \put(11,1){\circle{2}}
            \put(12,2){\makebox(1,1)[c]{$u$}}
            \put(12,1){\vector(1,0){3}}
            \put(15,0){\framebox(2,2)[c]{$b$}}
            \put(17,1){\vector(1,0){3}}
            \put(21,1){\circle{2}}
            \put(21,1){\circle*{0.6}}
            \put(22,2){\makebox(1,1)[c]{$v$}}
            \put(22,1){\vector(1,0){3}}
            \put(25,0){\framebox(2,2)[c]{$c$}}
            \put(27,1){\vector(1,0){3}}
            \put(31,1){\circle{2}}
            \put(32,2){\makebox(1,1)[c]{$f_1$}}
          \end{picture}
        }

        \put(-5,8){
          \setlength{\unitlength}{0.18cm}
          \begin{picture}(27,5)(-3,-1)
            \put(-3,-1){\framebox(27,5)[c]{}}
            \put(-2,2){\makebox(1,1)[c]{$\on{N}_2$}}
            \put(1,1){\circle{2}}
            \put(2,2){\makebox(1,1)[c]{$i_2$}}
            \put(2,1){\vector(1,0){3}}
            \put(5,0){\framebox(2,2)[c]{$d$}}
            \put(7,1){\vector(1,0){3}}
            \put(11,1){\circle{2}}
            \put(11,1){\circle*{0.6}}
            \put(12,2){\makebox(1,1)[c]{$s$}}
            \put(12,1){\vector(1,0){3}}
            \put(15,0){\framebox(2,2)[c]{$e$}}
            \put(17,1){\vector(1,0){3}}
            \put(21,1){\circle{2}}
            \put(22,2){\makebox(1,1)[c]{$f_2$}}
          \end{picture}
        }

        \put(18.6,4){
          \setlength{\unitlength}{0.16cm}
          \begin{picture}(46,12)
            \put(-1,-5){\framebox(44.5,13)[c]{}}
            \put(0.5,5){\makebox(1,1)[c]{$\on{N}_3$}}

            \put(5,8){\makebox(18,2)[c]{net-token produced on $\sn{r}$ by $\sn{t}$}}

            \put(1,1){\circle{2}}
            \put(2,2){\makebox(1,1)[c]{$i_3$}}
            \put(2,1){\vector(1,0){2}}
            \put(4,0){\framebox(2,2)[c]{$ $}}
            \put(6,2){\vector(1,1){2.3}}
            \put(6,0){\vector(1,-1){2.3}}

            \put(33.7,4.3){\vector(1,-1){2.3}}
            \put(33.7,-2.3){\vector(1,1){2.3}}
            \put(36,0){\framebox(2,2)[c]{$ $}}
            \put(38,1){\vector(1,0){2}}
            \put(41,1){\circle{2}}
            \put(42,2){\makebox(1,1)[c]{$f_3$}}

            \put(7.5,-4){
              \begin{picture}(32,4)
                \put(1,1){\circle{2}}
                \put(2,2){\makebox(1,1)[c]{$i_1$}}
                \put(2,1){\vector(1,0){2}}
                \put(4,0){\framebox(2,2)[c]{$a$}}
                \put(6,1){\vector(1,0){2}}
                \put(9,1){\circle{2}}
                \put(10,2){\makebox(1,1)[c]{$u$}}
                \put(10,1){\vector(1,0){2}}
                \put(12,0){\framebox(2,2)[c]{$b$}}
                \put(14,1){\vector(1,0){2}}
                \put(17,1){\circle{2}}
                \put(17,1){\circle*{0.6}}
                \put(18,2){\makebox(1,1)[c]{$v$}}
                \put(18,1){\vector(1,0){2}}
                \put(20,0){\framebox(2,2)[c]{$c$}}
                \put(22,1){\vector(1,0){2}}
                \put(25,1){\circle{2}}
                \put(26,2){\makebox(1,1)[c]{$f_1$}}
              \end{picture}
            }

            \put(7.5,4){
              \begin{picture}(32,4)
                \put(1,1){\circle{2}}
                \put(2,2){\makebox(1,1)[c]{$i_2$}}
                \put(2,1){\vector(1,0){4}}
                \put(6,0){\framebox(2,2)[c]{$d$}}
                \put(8,1){\vector(1,0){4}}
                \put(13,1){\circle{2}}
                \put(14,2){\makebox(1,1)[c]{$s$}}
                \put(13,1){\circle*{0.6}}
                \put(14,1){\vector(1,0){4}}
                \put(18,0){\framebox(2,2)[c]{$e$}}
                \put(20,1){\vector(1,0){4}}
                \put(25,1){\circle{2}}
                \put(26,2){\makebox(1,1)[c]{$f_2$}}
              \end{picture}
            }
          \end{picture}
        }

      \end{picture}
      \caption{\label{fig:hornet1} Modification of the Net-Token's Structure.}  
    \end{center}
    \vspace{-5mm}
  \end{figure}
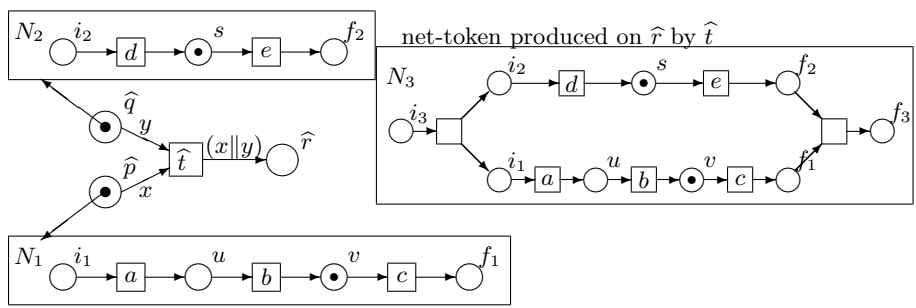

  \end{example}

\subsection{Events and Firing Rule}

\paragraph{Events}
The synchronisation labelling generates the set of system events
$\Theta$.  We have three kinds of events:

\begin{enumerate}
\item Synchronised firing: There is at least one object{} net that has
  to be synchronised, i.e.,  there is a $\on{N}$ such that $\sn{l}
  (\sn{t}) ( \on{N} )$ is not empty.

  Such an event is a pair $\theta = \sn{t}^{\alpha}[\vartheta]$, where
  $\sn{t}$ is a system net transition, $\alpha$ is a variable binding,
  and $\vartheta$ is a function that maps each object{} net to a
  multiset of its transitions, i.e.,  $\vartheta(\on{N}) \in
  \mathit{MS}(\on{T}_{\on{N}})$.
  It is required that $\sn{t}$ and $\vartheta(\on{N})$ have matching
  multisets of labels, i.e.,  \( \sn{l}(\sn{t})(\on{N}) =
  \on{l}_{\on{N}}^{\sharp} (\vartheta(\on{N})) \) for all $\on{N} \in
  \mathcal{U}$.
  (Remember that $\on{l}_{\on{N}}^{\sharp}$ denotes the multiset
  extension of $\on{l}_{\on{N}}$.)

  The intended meaning is that $\sn{t}$ fires synchronously with
  all the object{} net transitions $\vartheta(\on{N}), \on{N} \in
  \mathcal{U}$.

\item System-autonomous firing: The transition $\sn{t}$ of the system
  net fires autonomously, whenever $\sn{l} (\sn{t})$ is the empty
  multiset $\mathbf{0}$.

  We consider system-autonomous firing as a special case of
  synchronised firing generated by the function
  $\vartheta_{\mathit{id}}$, defined as $\vartheta_{\mathit{id}}
  (\on{N}) = \mathbf{0}$ for all $\on{N} \in \mathcal{U}$.

\item Object{} autonomous firing: An object{} net transition $\on{t}$ in
  $\on{N}$ fires autonomously, whenever $\on{l}_{\on{N}} (\on{t}) =
  \bot_{k}$.
  
  Object{} autonomous events are denoted as
  $\mathit{id}_{\sn{p},\on{N}}[\vartheta_{\on{t}}]$, where
  $\vartheta_{\on{t}} (\on{N}') = \{\on{t}\}$ if $\on{N} = \on{N}'$
  and $ \mathbf{0}$ otherwise.  The meaning is that in object net
  $\on{N}$ fires $\on{t}$ autonomously within the place $\sn{p}$.

  For the sake of uniformity we define for an arbitrary binding
  $\alpha$:
  \[
    \pre_{\alpha}(\mathit{id}_{\sn{p},\on{N}})(\sn{p}')(\on{N}') =
    \post_{\alpha}(\mathit{id}_{\sn{p},\on{N}})(\sn{p}')(\on{N}') =
    \begin{cases}
      1 & \text{ if } \sn{p}'= \sn{p} \land  \on{N}' =  \on{N}
      \\
      0 & \text{ otherwise. }
    \end{cases}
  \]
  We extend the guard function to these pseudo transitions by setting
  $\sn{G}(\mathit{id}_{\sn{p},\on{N}}) := \textsc{True}$.
\end{enumerate}

The set of all \emph{events} generated by the labelling $l$ is
$\Theta_l := \Theta_{1} \cup \Theta_{2}$, where $\Theta_{1}$ contains
synchronous events (including system-autonomous events as a special
case) and $\Theta_{2}$ contains the object{} autonomous events:

\begin{equation}
  \label{eq:sync-events}
  \begin{array}{rcll}   
    \Theta_{1} &:=&
                    \Big\{  \sn{\tau}^{\alpha}[\vartheta]  
    &\mid
      \forall   \on{N} \in \mathcal{U}: 
      \sn{l}_{\alpha}(\sn{t})(\on{N}) = \on{l}_{\on{N}}^{\sharp} (\vartheta(\on{N}))
      \Big\}  
    \\
    \Theta_{2} &:=&
                    \Big\{  
                    \mathit{id}_{\sn{p},\on{N}}[\vartheta_{\on{t}}]
    &\mid
      \sn{p} \in \sn{P},
      \on{N} \in \mathcal{U}_{\mathit{k}(\sn{p} )},
      \on{t} \in \on{T}_{\on{N}}
      \Big\}
  \end{array}
\end{equation}


\paragraph{Firing Rule}

A system event $\theta = \sn{\tau}^{\alpha}[\vartheta]$ removes net-tokens
together with their individual internal markings.  Firing the event
replaces a nested multiset $\lambda \in \mathcal{M}_H$ that is part of
the current marking $\mu$, i.e.,  $\lambda \sqsubseteq \mu$, by the
nested multiset $\rho$.
The enabling condition is expressed by the \key{enabling predicate}
$\phi_{\mathit{EH}}$ (or just $\phi$ whenever $\mathit{EH}$ is clear
from the context):

\begin{equation}
  \label{eq:firepredicate}  
  \begin{array}{rl}
    \phi_{\mathit{EH}}&(\sn{\tau}^{\alpha}[\vartheta], \lambda, \rho)
                        \iff
                        \forall k \in K:
    \\ &
         \forall \sn{p}\in  \mathit{k}^{-1}(k):
         \forall \on{N} \in \mathcal{U}_{k}: 
         \Pi^{{1}}_{\on{N}}(\lambda)(\sn{p}) = \pre_{\alpha}(\sn{\tau})(\sn{p})(\on{N}) \land{}
    \\ &
         \forall \sn{p}\in  \mathit{k}^{-1}(k):
         \forall \on{N} \in \mathcal{U}_{k}: 
         \Pi^{{1}}_{\on{N}}(\rho)(\sn{p}) = \post_{\alpha}(\sn{\tau})(\sn{p})(\on{N}) \land{}
    \\ &
         \Pi^{{2}}_{k}(\lambda)
         \geq
         \sum_{\on{N} \in \mathcal{U}_{k}}
         \pre_{\on{N}}^{\sharp}(\vartheta(\on{N})) \land{}
    \\ &
         \Pi^{{2}}_{k}(\rho) = \Pi^{{2}}_{k}(\lambda) 
         + \sum_{\on{N} \in \mathcal{U}_{k}}
         \post_{\on{N}}^{\sharp}(\vartheta(\on{N})) 
         - \pre_{\on{N}}^{\sharp}(\vartheta(\on{N})) 
  \end{array}
\end{equation}

The predicate $\phi_{\mathit{EH}}$ has the following meaning: 
\begin{itemize}
\item Conjunct (1) states that the removed sub-marking $\lambda$
  contains on $\sn{p}$ the right number of net-tokens, that are
  removed by $\sn{\tau}$.
\item   Conjunct (2) states that generated sub-marking $\rho$ contains on
  $\sn{p}$ the right number of net-tokens, that are generated by
  $\sn{\tau}$.
\item   Conjunct (3) states that the sub-marking $\lambda$ enables all
  synchronised transitions $\vartheta(\on{N})$ in the object $\on{N}$.
\item   Conjunct (4) states that the marking of each object net $\on{N}$ is
  changed according to the firing of the synchronised transitions
  $\vartheta(\on{N})$.
\end{itemize}

Note that conjuncts (1) and (2) assure that only net-tokens relevant
for the firing are included in $\lambda$ and $\rho$.  Conditions (3)
and (4) allow for additional tokens in the net-tokens.

For system-autonomous events
$\sn{t}^{\alpha}[\vartheta_{\mathit{id}}]$ the enabling predicate
$\phi_{\mathit{EH}}$ can be simplified further: Conjunct (3) is always true since
$\pre_{\on{N}}(\vartheta_{\mathit{id}}(\on{N})) = \mathbf{0}$.
Conjunct (4) simplifies to $ \Pi^{{2}}_{k}(\rho) = \Pi^{{2}}_{k}(\lambda)$,
which means that no token of the object nets get lost when a
system-autonomous events fires.

Analogously, for an object{} autonomous event $\sn{\tau}
[\vartheta_{\on{t}}]$ we have an idle-transition $\sn{\tau} =
\mathit{id}_{\sn{p},\on{N}}$ and $\vartheta = \vartheta_{\on{t}}$ for
some $\on{t}$.  Conjunct (1) and (2) simplify to
$\Pi^{{1}}_{\on{N'}}(\lambda) = \sn{p} = \Pi^{{1}}_{\on{N'}}(\rho) $ for
$\on{N'} = \on{N}$ and to $\Pi^{{1}}_{\on{N'}}(\lambda) = \mathbf{0} =
\Pi^{{1}}_{\on{N'}}(\rho) $ otherwise.  This means that $\lambda = \sn{p}[
\on{M}]$, $\on{M}$ enables $\on{t}$, and $\rho = \sn{p}[ \on{M} -
\pre_{\on{N}}(\sn{t}) + \post_{\on{N}}(\sn{t}) ]$.

\begin{definition}[Firing Rule]
  \label{def:ehornet-fire}
  Let $\mathit{EH}$ be an \eHornet{} and $\mu, \mu'
  \in \mathcal{M}_H$ markings.

  \begin{itemize}
  \item The event $\sn{\tau}^{\alpha}[\vartheta]$ is enabled in $\mu$
    for the mode $(\lambda, \rho) \in \mathcal{M}_H^{2}$ iff $\lambda
    \sqsubseteq \mu \land {\phi_{\mathit{EH}}}(\sn{\tau}[\vartheta], \lambda, \rho)$
    holds and the guard $\sn{G}(\sn{t})$ holds,
    i.e., $E  \models_{\mathcal{I}}^{\alpha} \sn{G}(\sn{\tau})$.

  \item An event $\sn{\tau}^{\alpha}[\vartheta]$ that is enabled in
    $\mu$ can fire -- denoted \( \mu
    \labtra{\sn{\tau}^{\alpha}[\vartheta](\lambda, \rho)}{\mathit{EH}}
    \mu' \).

  \item The resulting successor marking is defined as $\mu' = \mu -
    \lambda + \rho$.
  \end{itemize}

  Firing is extended to sequences $w \in \Theta_l^*$ in the usual way.
\end{definition}

Note that the firing rule has no a-priori decision how to distribute
the marking on the generated net-tokens. Therefore we need the mode
$(\lambda, \rho)$ to formulate the firing of
$\sn{\tau}^{\alpha}[\vartheta]$ in a functional way.

\begin{example}
  We will illustrate the firing rule considering the \Hornet{} from
  Fig.~\ref{fig:hornet1} again.
  To model a run-time adaption, we  combine $\on{N}_1$ and $\on{N}_2$
  resulting in the net $\on{N}_3 = (\on{N}_1 \| \on{N}_2)$.
  This modification is modelled by system net transition $\sn{t}$ of
  the \Hornet{}.
  In a binding $\alpha$ with $x \mapsto \on{N}_1$ and
  $y \mapsto \on{N}_2$ the transition $t$ is enabled.  Let
  $(x \| y)$ evaluate to $\on{N}_3$ for $\alpha$.
  When $\sn{t}$ fires it removes the two net-tokens from $\sn{p} $ and
  $\sn{q}$ and generates one new net-token on place $\sn{r}$.
  This is the event $\theta_1 = \sn{t}^{\alpha}[\vartheta]$, where
  $\vartheta = \vartheta_{\mathit{id}}$.
  The net-token generated on $\sn{r}$ has the structure of
  $\on{N}_3$ and its marking is obtained as a transfer from the
  token on $\on{v}$ in $\on{N}_1$ and the token on $\on{s}$ in
  $\on{N}_2$ into $\on{N}_3$:
  \[
    \sn{p}[\on{N}_1, \on{v}]
    +
    \sn{q}[\on{N}_2, \on{s}]
    \quad
    \labtr{\theta_1}
    \quad
    \sn{r}\big[\, (\on{N}_1\| \on{N}_2), \on{s} + \on{v}\big]
  \]
  This transfer is possible since all the places of $\on{N}_1$ and
  $\on{N}_2$ are also places in $\on{N}_3$ and tokens can be
  transferred in the obvious way.

  It is also possible that the net-tokens fire object autonomously.
  E.g., the net-token on place $\sn{q}$ enables the object net
  transition $\on{e}$ in $\mu_0$, i.e., the event
  $\theta_2 = \mathit{id}_{\sn{q},\on{N}_2}[\vartheta_{\on{e}}]$:
  \[
    \sn{p}[\on{N}_1, \on{v}]
    +
    \sn{q}[\on{N}_2, \on{s}]
    \quad
    \labtr{\theta_2}
    \quad
    \sn{p}[\on{N}_1, \on{v}]
    +
    \sn{q}[\on{N}_2, \on{f_2}]
  \]
  Analogously, the net-token on $\sn{p}$ enables the transition
  $\on{c}$.
\end{example}

\section{Stochastic Extensions of \eHornets{}}
\label{sec:stochastic-ehornets}

Let
$ \mathit{EH} = (\sn{N}, \mathcal{U}, \mathcal{I}, \mathit{k}, \Theta,
\mu_0) $ be an \eHornet{}.  The \key{reachability graph}
$\mathit{RG}(\mathit{EH}) = (V,E,\mu_0) $ contains all nested markings
$V= \mathit{RS}(\mathit{EH})$ as vertices (or nodes),
$E = \{ (\mu,\theta,\mu' ) \mid \mu\labtra{\theta}{} \mu' \}$ as edges
and the initial marking $\mu_0$ as a distinguished node.
We equip the \eHornet{ }model with a \key{rate} function
$\Lambda: \Theta \to \Real^{>0}$ that assigns firing rates to events
$\theta\in\Theta$.

\begin{definition}
A \key{stochastic \eHornet{}}
 $\mathit{SEH} = (\mathit{EH}, \Lambda) $
 is given by 
 an \eHornet{}
$ \mathit{EH} 
$ 
  and a firing rate 
  $\Lambda : \Theta \to \Real^{>0}$.
\end{definition}

\paragraph{Stochastic \eHornets{}: Event-Based Probabilities}

For Stochastic Petri nets (SPN) \cite{marsan1989} the usual way to
derive probabilities from these rates is to normalise over all
transitions enabled in a given marking.  In the following we extend
this idea to nested events of \eHornets{}.
Let
$\mathit{En}(\mu) := \{ \theta \in \Theta \mid \mu \labtr{\theta} \} $
be the set of all events enabled in the nested marking $\mu$.
Then, the probability of firing $\theta \in \mathit{En}(\mu) $ is
proportional to its rate $\Lambda(\theta) \in \Real^{>0}$.  For
\eHornets{} we define the firing probability as:
\begin{equation}
  \label{eq:100}
  \mathit{Pr}_{\mu}(\theta) :=
  \frac{  \Lambda(\theta) }{
    \sum_{\theta\in\mathit{En}(\mu) } \Lambda(\theta)}  
\end{equation}

For an arc $(\mu,\theta,\mu' )$ in the reachability graph
$\mathit{RG}(\mathit{EH}) = (V,E,\mu_0) $ we define
  \begin{equation}
  \mathit{Pr}_{}((\mu,\theta,\mu' ))
  :=
  \mathit{Pr}_{\mu}(\theta)
  \label{eq:500}  
\end{equation}
These probabilities turn the 
reachability graph into a (discrete) Markov chain.

\begin{definition}
 Let $\mathit{SEH} = (\mathit{EH}, \Lambda)$ be a  {stochastic \eHornet{}}.
 
 The induced \key{discrete Markov chain} is
 $\mathit{DMC}(\mathit{SEH}) = (V,E, \mathit{Pr},\mu_0)$, where
 $\mathit{RG}(\mathit{EH}) = (V,E,\mu_0) $ is the reachability graph
 of $\mathit{EH}$.
\end{definition}

The induced {discrete Markov chain} is defined in a `global' fashion
since it considers rates $\Lambda(\theta)$ of events and these events
are generated as a synchronisation of transitions from both, system-
and object-net.  In the following we like to derive these rates from
those of the transitions being involved in the event.

\paragraph{Stochastic \eHornets{}: Nested Firing Rates}

We can calculate the rates of an event by assigning rates to the
system net and to the object nets: Firing rates $\Lambda(\sn{t})$ for
the system net and $\Lambda(\on{t}) $ for the object nets.
Consider the event $\theta = \sn{\tau}^{\alpha}[\vartheta]$.  Here the
system net transition $\sn{\tau}$ is synchronised with a multiset of
transitions $\vartheta(\on{N})$ for each object net $\on{N}$.  (Here,
$\vartheta(\on{N}) = \mathbf{0}$ whenever the event does not
synchronise with the object net $\on{N}$.)

The rate of the event is generated from the system net and object net
rate done by the following \key{product rule} that multiplies all the
rates occuring in the event $\theta$ to describe the conjunction of
all transitions:

\begin{equation}
  \label{eq:4}  
  \Lambda_{pr}(\theta) :=
  \Lambda(\sn{\tau}) \cdot
  \prod_{k \in K}
  \prod_{ \on{N} \in \mathcal{U}_{k}}
  \prod_{t \in T_{\on{N}}}
  \Big(\Lambda_{\on{N}}(\on{t}) \Big)^{|\vartheta(\on{N})(\on{t})|}
\end{equation}


\begin{figure}[htbp]
  \centerline{\includegraphics[scale=0.36]{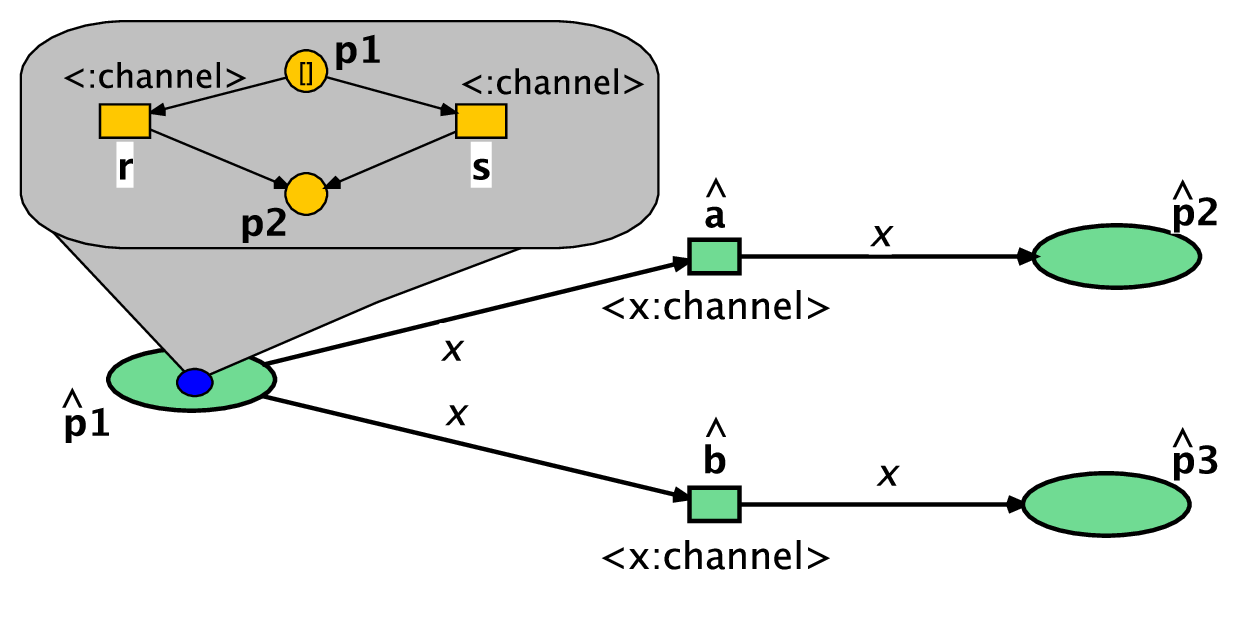}}
  \caption{\label{fig:scenario12/stoc-ehornet-example}{A Stochastic
      \eHornet{} }}
\end{figure}

\begin{example}
  For the example of
  Figure~\ref{fig:scenario12/stoc-ehornet-example}
  we have
  two system-net transition 
  $\sn{a}$
  and $\sn{b}$
  in conflict.
  They have to synchronise either with
  the object net transition $\on{r}$
  or $\on{s}$.
  From the perspective of synchronisation we have a symmetric situation here.
  We have the events:
  $
  \sn{a}[\on{N}\mapsto \on{r}]$,
  $\sn{b}[\on{N}\mapsto \on{s}]$,
  $\sn{b}[\on{N}\mapsto \on{r}]$, and
  $
  \sn{b}[\on{N}\mapsto \on{s}]
  $.

  Assume the following rates:
  $\Lambda( \sn{a}) = 2$,
  $\Lambda( \sn{b}) = 3$,
  $\Lambda( \on{r}) = 5$, 
  and
  $\Lambda( \on{r} ) = 7 $.
  From (\ref{eq:100}) we obtain the firing probability for e.g.
  $ \sn{a}[\on{N}\mapsto \on{r}] $ as
  \[
    \begin{array}{rcl}
      \mathit{Pr}\left( \sn{a}[\on{N}\mapsto \on{r} ]\right)
      &=&
          \frac{
          \Lambda(\sn{a})
          \cdot 
          \Lambda(\on{r})
          }
          {
          \Lambda(\sn{a})
          \cdot 
          \Lambda(\on{r})
          +
          \Lambda(\sn{a})
          \cdot 
          \Lambda(\on{s})
          +
          \Lambda(\sn{b})
          \cdot 
          \Lambda(\on{r})
          +
          \Lambda(\sn{b})
          \cdot 
          \Lambda(\on{s})
          }
      \\[2mm]
      &=&
          \frac{
          2 \cdot 5
          }{
          2 \cdot 5
          +
          2 \cdot 7
          +
          3 \cdot 5
          +
          3 \cdot 7
          }
          =
          \frac{10}{60} = 
          0,166..
    \end{array}
  \]
  The symmetry in the synchronisation leads to a symmetry in
  probabilities; e.g.; the probability to fire the system-net
  transition $\sn{a}$ with any possible object-net transition is:
  \[
    \begin{array}{rcccl}
      &&
         \frac{
         \Lambda(\sn{a})
         \cdot 
         \Lambda(\on{r})
         +
         \Lambda(\sn{a})
         \cdot 
         \Lambda(\on{s})
         }
         {
         \Lambda(\sn{a})
         \cdot 
         \Lambda(\on{r})
         +
         \Lambda(\sn{a})
         \cdot 
         \Lambda(\on{s})
         +
         \Lambda(\sn{b})
         \cdot 
         \Lambda(\on{r})
         +
         \Lambda(\sn{b})
         \cdot 
         \Lambda(\on{s})
         }
      &&
      \\[3mm]
      &=&
          \frac{
          \Lambda(\sn{a})
          \cdot 
          (\Lambda(\on{r})
          +
          \Lambda(\on{s})
          )
          }
          {
          \Lambda(\sn{a})
          \cdot 
          (
          \Lambda(\on{r})
          +
          \Lambda(\on{s})
          )
          +
          \Lambda(\sn{b})
          \cdot 
          (
          \Lambda(\on{r})
          +
          \Lambda(\on{s})
          )
          }
      &=&
          \frac{
          \Lambda(\sn{a})
          }
          {
          \Lambda(\sn{a})
          +
          \Lambda(\sn{b})
          }
    \end{array}
  \]
  This is the firing probability one would expect when considering the
  system-net alone.

  We have an analogous situation for the object.  Let us consider the
  probability of firing an event containing the object-net transition
  $\on{r}$:
  \[
    \begin{array}{rcccl}
      &&
         \frac{
         \Lambda(\sn{a})
         \cdot 
         \Lambda(\on{r})
         +
         \Lambda(\sn{b})
         \cdot 
         \Lambda(\on{r})
         }
         {
         \Lambda(\sn{a})
         \cdot 
         \Lambda(\on{r})
         +
         \Lambda(\sn{a})
         \cdot 
         \Lambda(\on{s})
         +
         \Lambda(\sn{b})
         \cdot 
         \Lambda(\on{r})
         +
         \Lambda(\sn{b})
         \cdot 
         \Lambda(\on{s})
         }
      &&
      \\[3mm]
      &=&
          \frac{
          (
          \Lambda(\sn{a})
          +
          \Lambda(\sn{b})
          )
          \cdot 
          \Lambda(\on{r})
          }
          {
          (
          \Lambda(\sn{a})
          +
          \Lambda(\sn{b})
          )
          \cdot 
          \Lambda(\on{r})
          +
          (
          \Lambda(\sn{a})
          +
          \Lambda(\sn{b})
          )
          \cdot 
          \Lambda(\on{s})
          }
      &=&
          \frac{
          \Lambda(\on{r})
          }
          {
          \Lambda(\on{r})
          +
          \Lambda(\on{s})
          }
    \end{array}
  \]
  This is the firing probability when considering the object-net
  alone.

  Note, that this effect is due to the symmetry in synchronisation;
  whenever the system-net transitions would have different
  synchronisation partners the probabilities would be different.
  Assume that e.g. $\sn{b}$ has much more synchronisation partners
  than $\sn{a}$, then the probability of events containing $\sn{a}$
  would have been dominated by the events containing $\sn{b}$.
\end{example}

\section{Modeling Self-Adaptive Systems}
\label{sec:stochastic-ehornets-for-mape}

Usually, \eHornets{} are used to specify self-modifying systems.  In
the following we like to show that for these scenarios the firing
rates can be naturally derived, i.e., models of self-adaptive systems
based on \eHornets{} can be extended to Stochastic \eHornets{} in an
automated way

In the following, we consider a very common scenario, where the
system-net describes the MAPE-loop of adaption
(monitor-analyse-plan-execute \cite{Weyns2020}), while the object-nets
are workflow nets \cite{Aalst97} composed by basic process algebraic
operations, like sequence, and-composition, and xor-choices.
For workflow nets we have have natural candidates for the rates
$\Lambda(\on{t})$ in the object-nets -- we usually derive them from
execution logs, i.e. by monitoring data.  Here, rates are used to
describe probabilities of xor-branches.

Note, that for the object-nets there is no need for an additional
concept as we can easily integrate the rates into the structure of the
net-token: Object Nets are pairs of the net topology $\on{N}$ and the
firing rate $\Lambda $, denoted $\on{N}^{\Lambda}$.  The mapping
$\Lambda$ assigns firing rates to transitions.  Therefore, we have an
independent rate for each net-token $[\on{N}^{\Lambda}, \on{m}]$.

When we have self-modifying systems, the transitions $\sn{t}$ in the
system-net describe transformations of these workflows.  The rates
$\Lambda(\sn{t})$ should describe the probability of executing the
transformation that is described by the system-net transition $\sn{t}$
during the MAPE-loop.
When considering self-adaptive systems we like to express that there
is an relationship between the complexity of an adaption and its
probability.  The motivations for this is that transformations have to
be evaluated during some kind of planning process and transformations
that are less complex are usually more frequently considered during
this planning.

For \eHornets{} we have natural candidates that describe the
\key{transformation complexity}: the guard $\sn{G}(\sn{t})$ and the
arc inscriptions $\pre(\sn{t})(\sn{p})$ and $ \post(\sn{t})(\sn{p})$.
The transformation complexity of a system-net transition $\sn{t}$ is
the number of operations $\| \cdot \|$ occuring in its arc
inscriptions and guards:
\begin{equation}
    \mathit{TC}(\sn{t})
    :=
    \|\sn{G}(\sn{t})\|
    +
    \sum_{\sn{p} \in \sn{P}}
    \|\pre(\sn{t})(\sn{p}))\| 
    +
    \sum_{\sn{p} \in \sn{P}}
    \|\post(\sn{t})(\sn{p}))\|  
\end{equation}

The idea to derive the firing rate from $\mathit{TC}(\sn{t})$ is
straightforward: Since the search space for planning grows
exponentially in the number of transformation operators, we define
that the rates drop exponentially, too:
\begin{equation}
    \Lambda_{mape}(\sn{t})
    :=
    \gamma^{ \mathit{TC}(\sn{t})},
    \qquad
    \gamma \in [0;1]
    \label{eq:200}
\end{equation}

Here, $\gamma$ is a meta-parameter that specifies the discount for the planning horizon.
Note, in this MAPE-loop setting the firing rates are directly derived
from the model without any need for extra modeling effort.  Thus,
(\ref{eq:200}) turns an qualitative \eHornet{} into a quantitative
stochastic model -- and this for free.

\section{Example: Modelling Adaption in a Coordination Game}
 \label{sec:stoch-ehornet-bos}

 The following example for a self-adaptive system is based on the
 battle-of-sexes scenario, which is well-known in game theory.
 Two agents, named $0$ and $1$, must choose between two actions,
 labelled as $a_i$ and $b_i$, $i=0,1$. They receive a positive reward
 if they choose the same action and zero otherwise.

\begin{flushleft}
    \parbox{0.73\textwidth}{
      In this game, the first agent prefers action $a$, while the
      second prefers $b$.
      If we assume that the reward for the preferred outcome is three
      times higher than for the other, then the game is specified by
      the following payoff matrix.  }
    \;  
    \(
    \begin{array}{c||c|c|}
         & a_1 & b_1 \\ \hline
         \hline
       a_0 \; & \;(3,1) \;& \;(0,0) \;\\ \hline
     b_0\;   & \;(0,0)\; & \;   (1,3)\;\\ \hline
    \end{array}
  \)     
\end{flushleft}

\begin{figure}[htbp]
    \centerline{\includegraphics[width=\textwidth]{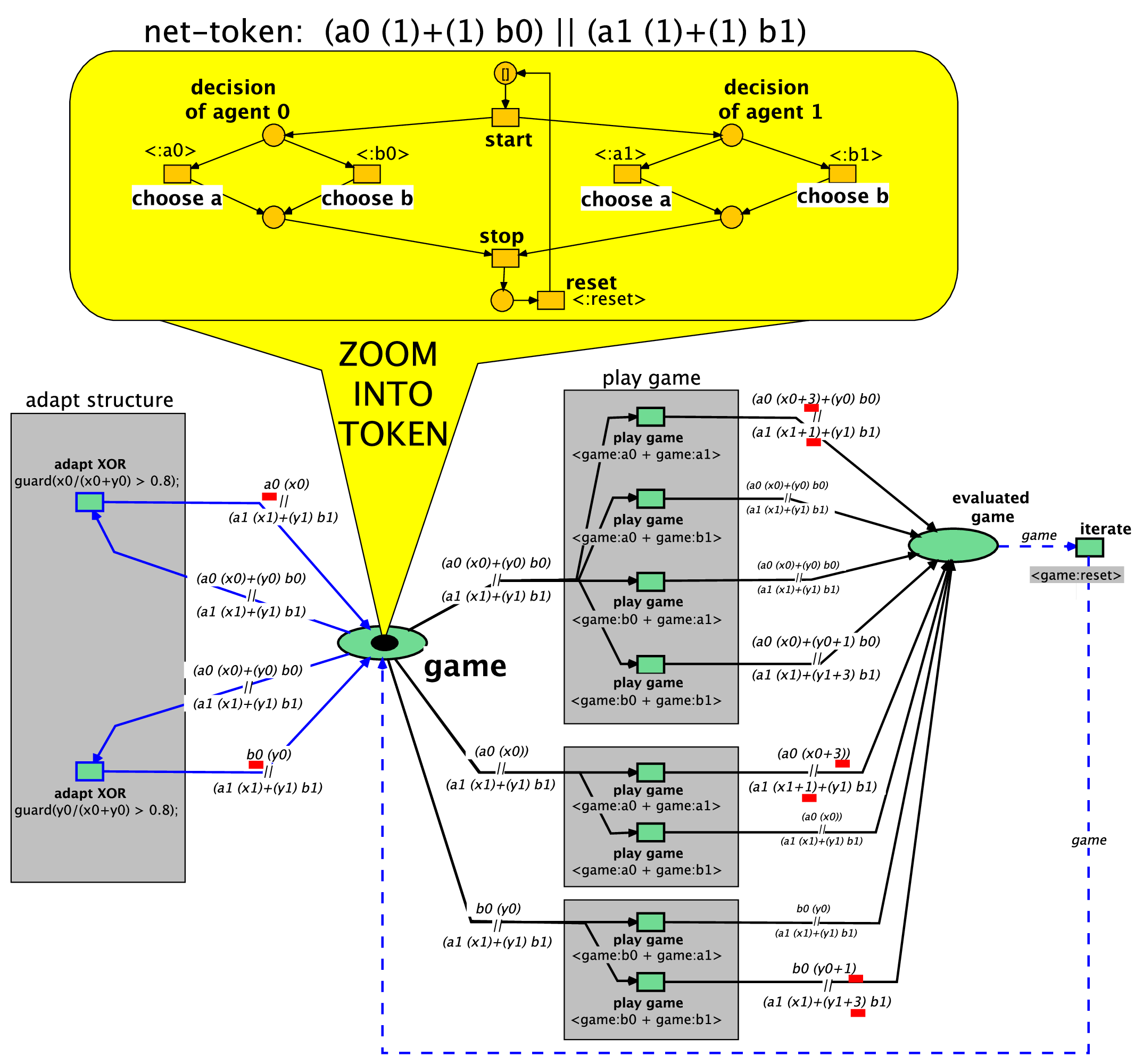}}
    \caption{\label{fig:scenario12/mape-ehornet}{The
        \eHornet{}: System-Net containing the Battle-of-Sexes
        Interaction (right)     and the Structural Adaption Logic (left)}}
  \end{figure}

Let $(a \prXOR{A}{B} b) $ describe the xor-choice between action $a$
and $b$ where $\Lambda(a) = A$ and $\Lambda(b) = B$.
The object net that models this game is shown as a net-token in
Fig.~\ref{fig:scenario12/mape-ehornet}; it is a parallel composition
(denoted by
$\_ \| \_$)
of two choices:
\[
\on{N}_1^{\Lambda}
=
(a_0 
\prXOR{A_0}{B_0} b_0) \,\,\|\,\,
(a_1 \prXOR{A_1}{B_1} b_1)
\]

The system net observes the decision history and adapts by modifying
the rates (cf. the \eHornet{} in
Fig.~\ref{fig:scenario12/mape-ehornet}).  We have four transitions
named \textsf{play game} on the right side corresponding to the four
different ways of choosing the actions.  We give the payoff as a
reward signal to the agents. (There might be more appropriate ways of
adapting, but for this simple example we do not care about the
efficiency of the learning process.)  For example, when the agents
play $(a_0,a_1)$ then we update the rates in the the workflow by the
payoff $(3,1)$ and we obtain:
\[
\on{N}_2^{\Lambda}
=
(a_0  \prXOR{A_0+3}{B_0} b_0) \,\,\|\,\,
(a_1 \prXOR{A_1+1}{B_1} b_1)
\]


We have another source of adaption in the system net: Choices which
are chosen quite regularly over a longer time period are converted
into fixed structures without choice by the two transitions named
\textsf{adapt XOR} on the left-hand side.  In this example the
transformation is allowed whenever $a_0$ is chosen in more than 80\%
of the time.  This is expressed by the transition guard
$ \frac{x_0}{(y_0+y_0) }> 0.8 $.
Then, we obtain \( a_0 \,\|\, (a_1 \prXOR{A_1}{B_1} b_1) \) as the
modified net structure.  Analogously whenever $b_0$ dominates.  (For
simplicity we omit modifications whenever the second agent has a
dominating option.)

  \begin{figure}[thbp]
    \centerline{\includegraphics[width=1.0\textwidth]{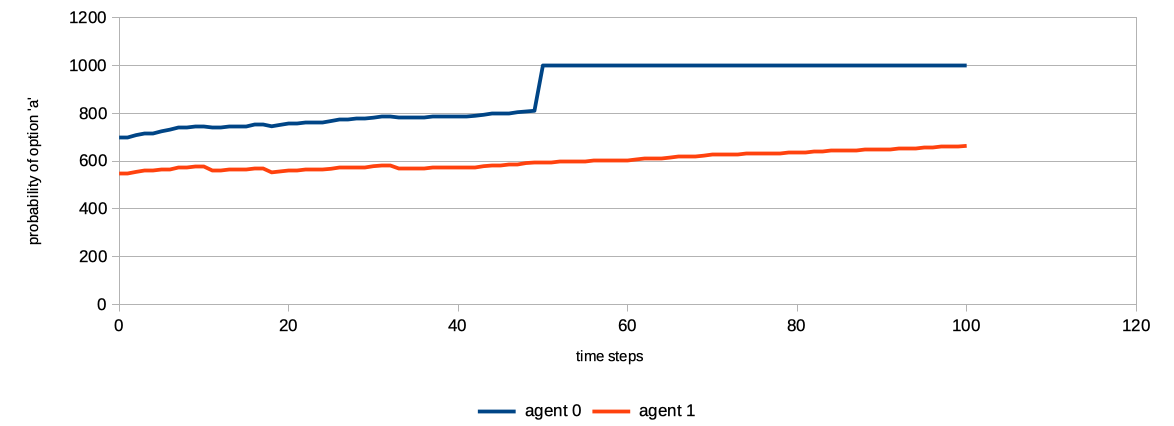}}
    \caption{\label{fig:scenario12/log105}{A Sample Run: The dynamics of the Probabilities of  Options $a_{0,1}$ }}
  \end{figure}


For an example run 
(with initial rates 
$A_0 = 70.0$,
$B_0=30.0$,
$A_1 = 55.0$, and
$B_1 = 45.0$)
the probabilities of choosing option $a$ is given in Fig.~\ref{fig:scenario12/log105}.
We choose $\gamma = 0.5$ to balance the frequencies of structural
modifications and rate updates.
One can clearly observe that the update rule increases the
probabilities in favor of options $a_0$ and $a_1$.  Note, the `jump'
at time $t=50$: Here, a structural modification takes place, which
sets the probability of choosing $a_0$ to $100\%$.

\section{Conclusion  }

In this contribution we introduced \emph{Stochastic \eHornets{}}, a
Nets-within-Nets formalism, where the system net and each net-token is
equipped with firing rates.
Our formalism is well-suited to express quantitative aspects for
self-adaptive systems.  It is a typical setting that the system-net
describes the adaption loop and the net-tokens describe some kind of
process logic (e.g., protocols or workflow nets) that are subject to
modification at run-time.  For these and similar scenarios the firing
rates arise naturally from the application domain: The workflows
contain rates for xor-choices and for adaption events in the system
net we argued that the rate is inversely proportional to the
transformation complexity $\mathit{TC}(\sn{t})$.


In ongoing work we will deepen the aspect of formal analysis for these
stochastic \eHornets{}.
A major challenge is that the state space of an \eHornet{} grows much
faster than that of a p/t net: The reachability problem needs double
exponential time even for safe \eHornets{}
\cite{koehler+13-atpn,Koehler13-fi-hornets}.
At the moment we see at least two candidates for the analysis: First,
we will use Maude and its in-build stochastic features
(cf. \cite{journals/entcs/AghaMS06}) to express probabilities.  This
will allow us to use the state space exploration techniques as
provided by Maude (which we already used for \eHornets{} without
firing rates \cite{enase23,Capra+2024-tcs}).  Another candidate for
analysis is the translation into GreatSPN \cite{modelling-gspn-95},
which offers powerful techniques; a major challenge here is the
question how we can translate the algebraic structure of \eHornets{}
into a Stochastic Petri Net.

Abstractions are a complementary approach to tackle the
double-exponentially growing state space of an \eHornet{}: We already
know that \eHornet{} firing is preserved by projection onto the system
net.
Therefore, we like to study to which extend we can analyse our model
when considering the -- much smaller -- projection of the state space
alone.
Additionally, we like to exploit symmetries in the structure of an
\eHornet{} using the concept of automorphism (cf. our previous
definition of \EOS{}-automorphisms
\cite{CapraKoehler2024-EOS-morphism}).
%

\bibliographystyle{splncs04}
\bibliography{mk/defs,mk/koehler,mk/agent,mk/new,mk/eigenes,mk/konferenzen}

\end{document}